\documentclass[12pt]{iopart}
\usepackage{amssymb}
\usepackage{graphicx}
\usepackage{dcolumn}
\usepackage{bm}

\newcommand{\om}{\omega}
\newcommand{\vfi}{\varphi}
\newcommand{\bfr}{{\bf r}}
\newcommand{\prt}{\partial}

\newcommand{\bk}{{\bf{k}}}

\begin{document}

\title[Matter sound waves in two-component Bose-Einstein condensates]
{Matter sound waves in two-component Bose-Einstein condensates}
\author{B B Baizakov$^1$, A M Kamchatnov$^2$ and M Salerno$^3$}
\address{$^1$ Physical - Technical Institute, Uzbek Academy of
Sciences, 2-b, G. Mavlyanov str., 100084, Tashkent, Uzbekistan \\
$^2$ Institute of Spectroscopy, Russian Academy of Sciences, Troitsk,
Moscow Region, 142190, Russia \\
$^3$ Dipartimento di Fisica ``E. R. Caianiello'', Consorzio
Nazionale Interuniversitario per le Scienze Fisiche della Materia
(CNISM), Universit\'a di Salerno, via S. Allende I-84081,
Baronissi (SA), Italy } \ead{baizakov@uzsci.net,
kamch@isan.troitsk.ru, salerno@sa.infn.it}

\begin{abstract}

The creation and propagation of sound waves in two-component
Bose-Einstein condensates (BEC) are investigated and a new method
of wave generation in binary BEC mixtures is proposed. The method
is based on a fast change of the inter-species interaction
constant and is illustrated for two experimental settings: a
drop-like condensate immersed into a second large repulsive
condensate, and a binary mixture of two homogeneous repulsive
BEC's. A mathematical model based on the linearized coupled
Gross-Pitaevskii equations is developed and explicit formulae for
the space and time dependence of sound waves are provided.
Comparison of the analytical and numerical results shows excellent
agreement, confirming the validity of the proposed approach.

\end{abstract}
\pacs{03.75.Kk }
\submitto{\JPB}
\maketitle

\section{Introduction}

Wave phenomena in Bose-Einstein condensates (BEC) represent a very
active field of research since the first experimental realization
of BEC. Matter wave solitons, shock waves and different wave
instabilities are among the hot topics of the field presently
under investigation (for a recent review see \cite{carretero}). In
particular, dispersive shock waves in BECs consisting of modulated
nonlinear periodic waves which decompose into dark soliton trains
(e.g., when an initial disturbance is represented by a hole in the
density distribution) or eventually into small amplitude waves
(e.g., when an initial disturbance is represented by a hump in the
density distribution), have been recently investigated. Theory of
such waves and soliton trains described by the nonlinear
Schr\"odinger (NLS) equation  was developed in
\cite{gk-87,eggk-95,kku-2002} and applied to their formation in a
one-component BEC in \cite{bk-2003,kgk-2004,hoefer}. Shock waves
in discrete NLS models were investigated in \cite{KS97,ABDKS01}
while experimental observations of shock waves in BEC were
reported in \cite{hoefer,simula}. Although these studies were
mainly concentrated on single component BEC, binary BEC mixtures,
both single species (two distinct spin or hyperfine states of the
same atom) \cite{single} and double species \cite{double}, have
been created and the existence of several kinds of nonlinear
excitations in presence of optical lattices has been demonstrated
\cite{binary-mixtures}.

In spite of the intrinsic nonlinearity of BECs and the relevance
of nonlinear excitations for applications, small amplitude
wave-packets associated with long-wavelength modulations of the
condensate (matter sound waves) also appear to be important. These
waves naturally arise in the linear or quasi-linear regimes, e.g.
when the propagation of small amplitude disturbances on top of the
condensate is considered or when the inter-particle interactions
(nonlinearity) is ``artificially" reduced by means of Feshbach
resonances. They are generated also in a supersonic flow of BEC
past an obstacle where they form so called ``ship wave'' patterns
located outside the Mach cone
\cite{carusotto-2006,gegk-2007,gk-2007,gsk-2008}. Small amplitude
wave-packets should not be confused with dispersive shocks for
which larger disturbances or larger nonlinearity are usually
required (this is particularly true in the multi-dimensional
case). From a mathematical point of view they can be characterized
as waves which, opposite to solitons, entirely belong to the
continuum spectrum of the linear eigenvalue problem associated
with the NLS equation. From a physical point of view, these waves
are composed of linear modes which have a Bogoliubov-like
dispersion relation and which, in a pure quantum description,
correspond to elementary excitations of the system. The spectrum
of such excitations characterizes the properties of the
two-component BEC superfluid \cite{goldstein}.

Propagation of sound waves in single-component harmonically
trapped BEC's was experimentally observed in
\cite{simula,andrews97} in the form of travelling density
perturbations. Tightly focused far blue-detuned from atomic
transitions laser beam applied to the condensate acts as expulsive
potential and rapidly pushes atoms from the light field maximum.
Depending on the laser power and pulse duration, different kinds
of waves are created in the condensate. Smoothly propagating sound
waves are created when the laser beam waist is narrow compared to
the size of the condensate and the light intensity is small
enough. Sound waves in two-component BEC's were experimentally
observed in \cite{dutton}. While the advantages of experimental
techniques for BECs involving laser beams are extremely appealing,
there is a crucial drawback when it concerns applications to two
component condensates (in Ref. \cite{dutton} the slow light
technique was used). The reason for this is connected to the fact
that different atomic species feel different light-induced
expulsive potentials (due to optical dipole force) and respond to
the laser field in different manners. This leads to the appearance
of interfering waves which complicate very much the process under
investigation. From this point of view the search of alternative
methods to generate matter waves in two component BECs is highly
desirable.

The aim of this paper is twofold. From one side we investigate,
both analytically and numerically, sound waves in two-component
BEC's  generated by the time evolution of a small initial
disturbance. From the other side, we present a new method for the
generation of different kinds of waves in binary BEC mixtures
which is  based on the fast change of the inter-species
interaction constant. In view of the recent progress made in the
manipulation of the inter-component interactions in binary BEC
mixtures \cite{papp} by means of Feshbach resonances, the proposed
method appears to be experimentally quite feasible and has the
advantage that it relies entirely on the internal physical
mechanism for the creation of the density disturbance. This method
is used to investigate sound waves in two experimental settings:
i) the case of a drop-like condensate immersed into a second large
repulsive condensate and ii) the case of a homogeneous mixture of
two repulsive BEC's both with finite backgrounds. In case i) we
first create an initial disturbance in the form of a gray or a
bright hump (depending on the sign of the interaction) on a finite
background by adiabatically turning on the interaction between the
drop-like condensate and the surrounding condensate, and then we
rapidly remove the inter-species interaction. The energy is then
entirely  released in the form of sound waves in the background
condensate. In case ii) the initial disturbance is created by a
change of the intra- or inter-species interaction constant via
optically induced Feshbach resonance. To illustrate our approach
we numerically simulate the creation and evolution of matter sound
waves in a two component BEC. A mathematical model based on the
linearized Gross-Pitaevskii equation is has been developed and
explicit formulae for the space and time dependence of the sound
waves are obtained. Comparison of the analytical and numerical
results shows a very good agreement, confirming the validity of
the proposed approach. The obtained results can be of interest for
the physics of interacting two-component superfluid systems.

The outline of the paper is as follows. In Sec.~II we illustrate
the proposed approach for the one dimensional setting, by
formulating the mathematical model and performing numerical
simulations. In this case one component is a bright soliton or
drop-like condensate immersed in the second large repulsive
condensate with a finite background. Since the first component is
on the zero background, after the interspecies interaction is
turned off, we consider the waves generated in the second
(background) condensate. Analytical formulae describing the
space-time patterns of sound waves in coupled BECs are explicitly
derived. In Sec.~III the proposed method is extended to the two
dimensional case. In Sec.~IV we consider the most interesting and
experimentally relevant case of sound waves of two component
repulsive condensates both with a finite background. In this case
the inter-species interaction is present during the whole
evolution of waves in both condensates, exhibiting interference
effects in the dynamics. In Sec.~V the main results are
summarized.

\section{One-dimensional condensates}

\subsection{Decay of a small disturbance of a stationary state
to linear wave packets}

\begin{figure}[htb]
\centerline{
\includegraphics[width=6cm, height=4cm]{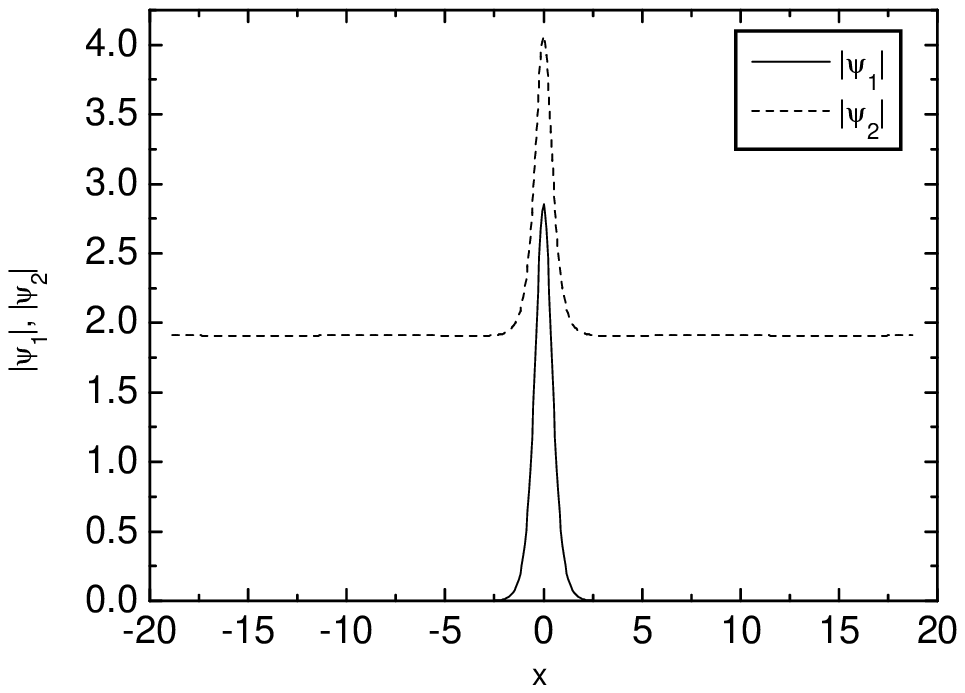} \qquad
\includegraphics[width=6cm, height=4cm]{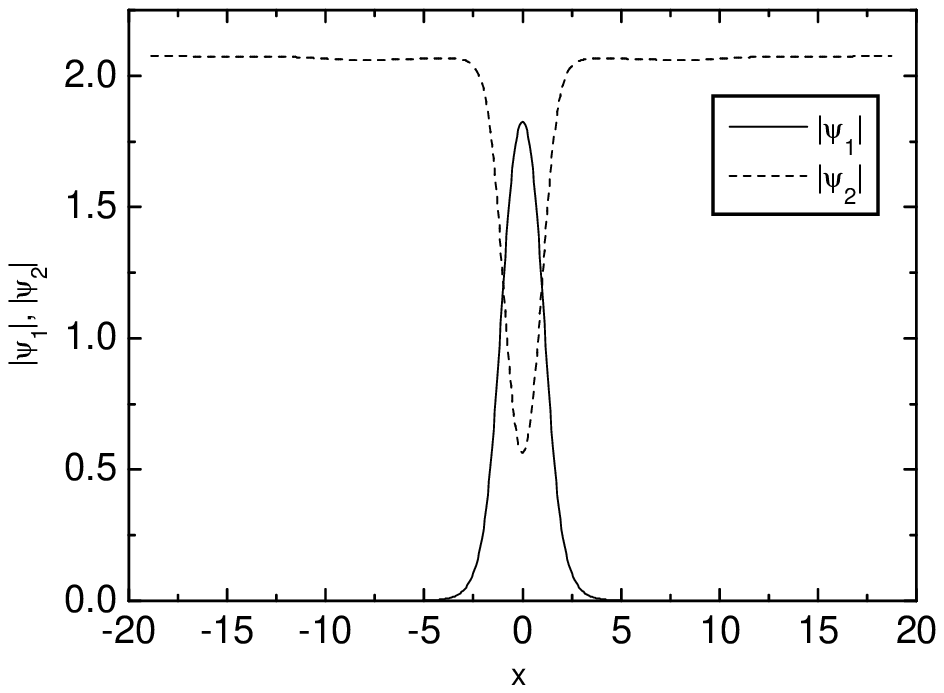}}
\caption{The stationary state of two coupled 1D BECs emerged from
the attractive (left panel), and repulsive (right panel)
inter-species interactions. Zero background condensate (solid
line) is attractive, and finite background condensate (dashed
line) is repulsive. } \label{fig1}
\end{figure}

Consider a quasi-1D two-component BEC with attractive interaction
in one species and repulsive one in the other species. In absence
of the interspecies interaction the stationary ground state
corresponds to a bright soliton in the attractive condensate and
the Thomas-Fermi (TF) distribution in the repulsive one (which can
be considered as uniform if the TF radius is much greater than the
bright soliton's width). Adiabatic turning on the interspecies
interaction leads to the formation of a new stationary state with
a hump or a hole in the repulsive BEC, depending on the sign of
the interspecies interaction (see Fig.~\ref{fig1}).

A fast turning off of the interspecies interaction will then
create a disturbed stationary state in the repulsive condensate.
In this section, we shall study evolution of such a perturbation.
Since the attractive condensate has a zero background density, we
can consider only waves created in the repulsive finite background
condensate using the scalar GP equation.

In this case dynamics of the repulsive BEC is governed by the
GP equation
\begin{equation}\label{2-1}
    i\psi_t+\frac12\psi_{xx}+(\mu-g|\psi|^2)\psi=0,
\end{equation}
where
\begin{equation}\label{2-2}
    \mu=g\rho_0=c^2
\end{equation}
is the chemical potential, $\rho_0$ is the density of an
undisturbed uniform state, and $c$ denotes the sound velocity of
long wavelength waves propagation along such a uniform state. We
imply here that the coupling constant $g\equiv g_{1D}$ corresponds
to the 1D geometry of the trap with strong radial confinement
($\omega_{\bot} \gg \omega_x$) of the BEC.

Let the initial state be described by a disturbed wave function
\begin{equation}\label{2-3}
    \psi_0=\sqrt{\rho_0}+\delta\psi_0,\quad |\delta\psi_0|\ll \sqrt{\rho_0}.
\end{equation}
For example, $\delta\psi_0$ can describe the hump or the hole
created in the density distribution by the interaction with
another condensate. Evolution of a small disturbance $\delta\psi$
is governed by a linearized equation (\ref{2-1}),
\begin{equation}\label{2-4}
    i\delta\psi_t+\frac12\delta\psi_{xx}-c^2(\delta\psi+\delta\psi^*)=0,
\end{equation}
which should be solved with the initial condition
\begin{equation}\label{2-5}
    \left.\delta\psi\right|_{t=0}=\delta\psi_0(x).
\end{equation}
It is convenient to separate in this equation the real and
imaginary parts by introducing
\begin{equation}\label{2-6}
    \delta\psi=A+iB,
\end{equation}
so that it transforms to the system
\begin{equation}\label{2-7}
    \begin{array}{l}
     A_t+\frac12B_{xx}=0,\\
     B_t-\frac12A_{xx}+2c^2A=0.
    \end{array}
\end{equation}
The variable $B$ can be easily excluded and we arrive at a single linear
equation
\begin{equation}\label{2-8}
    A_{tt}-c^2A_{xx}+\frac14A_{xxxx}=0,
\end{equation}
which can be readily solved by the Fourier method. To this end, we
notice that linear harmonic waves $A\propto\exp[i(kx-\om t)]$
satisfy, as it should be, to the Bogoliubov dispersion law,
\begin{equation}\label{2-9}
    \om=\pm\om(k),\quad \om(k)=k\sqrt{c^2+\frac{k^2}4}.
\end{equation}
Hence the general solution of Eq.~(\ref{2-8}) has the form
\begin{equation}\label{2-10}
    A(x,t)=\int_{-\infty}^{\infty}W_1(k)e^{i(kx-\om(k)t)}\frac{dk}{2\pi}+
    \int_{-\infty}^{\infty}W_2(k)e^{i(kx+\om(k)t)}\frac{dk}{2\pi},
\end{equation}
where $W_{1,2}(k)$ are to be determined from the initial conditions:
\begin{equation}\label{2-11}
    \begin{array}{l}
     A(x,0)\equiv A_0(x)=\int_{-\infty}^{\infty}[W_1(k)+W_2(k)]e^{ikx}\frac{dk}{2\pi},\\
     A_t(x,0)\equiv A_{t0}(x)=-i\int_{-\infty}^{\infty}\om(k)[W_1(k)-W_2(k)]e^{ikx}
    \frac{dk}{2\pi}.
    \end{array}
\end{equation}
The inverse Fourier transform gives the equations
$$
A_0(k)=W_1(k)+W_2(k),\quad A_{t0}(k)=-i\om(k)[W_1(k)-W_2(k)],
$$
which readily provide
\begin{equation}\label{2-12}
\begin{array}{l}
    W_1(k)=\frac12\left(A_0(k)+\frac{i}{\om(k)}A_{t0}(k)\right),\\
    W_2(k)=\frac12\left(A_0(k)-\frac{i}{\om(k)}A_{t0}(k)\right).
    \end{array}
\end{equation}
The first equation (\ref{2-7}) allows to express $A_{t0}(k)$ in
terms of $B_0(k)$, $A_{t0}(k)=(k^2/2)B_0(k)$. At last, $A(k)$ and
$B(k)$ can be related with disturbances of the density and the
phase in BEC,
$\psi=\sqrt{\rho_0+\delta\rho}\exp(i\delta\phi)\cong\sqrt{\rho_0}+
\frac{\delta\rho}{2\sqrt{\rho_0}}+i\sqrt{\rho_0}\delta\phi$, so
that
\begin{equation}\label{2-13}
    A=\frac{\delta\rho}{2\sqrt{\rho_0}},\quad B=\sqrt{\rho_0}\delta\phi.
\end{equation}
Hence, Eqs.~(\ref{2-12}) take the form
\begin{equation}\label{2-14}
\begin{array}{l}
    W_1(k)=\frac1{4\sqrt{\rho_0}}\left(\delta\rho_0(k)+i\frac{k^2\rho_0}{\om(k)}
    \delta\phi_0(k)\right),\\
    W_2(k)=\frac1{4\sqrt{\rho_0}}\left(\delta\rho_0(k)-i\frac{k^2\rho_0}{\om(k)}
    \delta\phi_0(k)\right).
    \end{array}
\end{equation}

\begin{figure}[htb]
\centerline{
\includegraphics[width=6cm, height=4cm,clip]{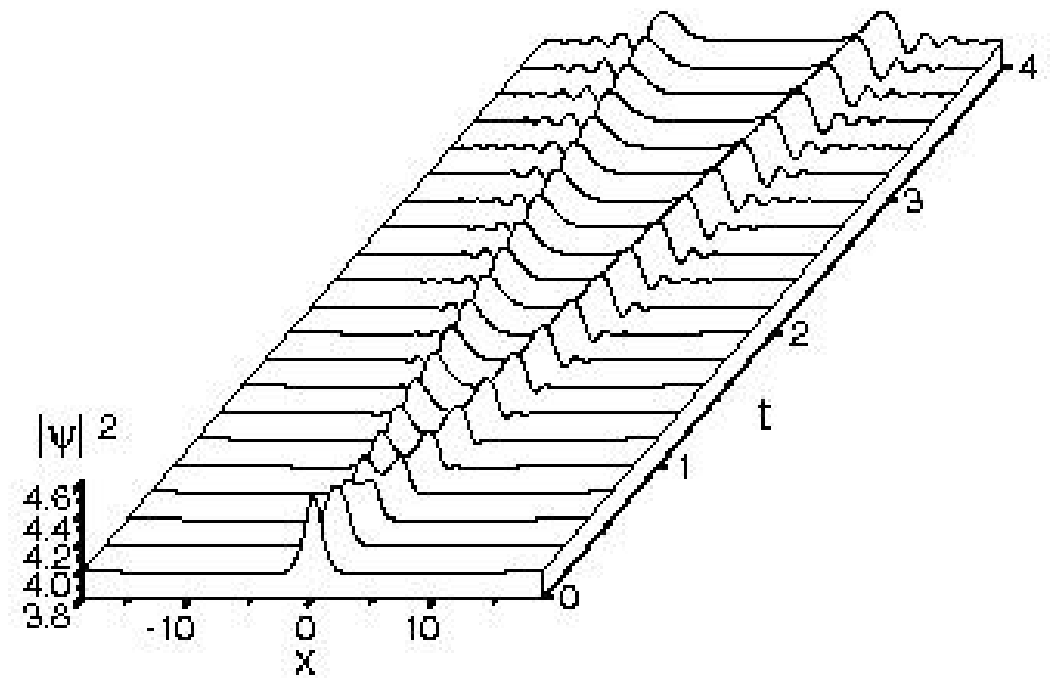}
\includegraphics[width=6cm, height=4cm]{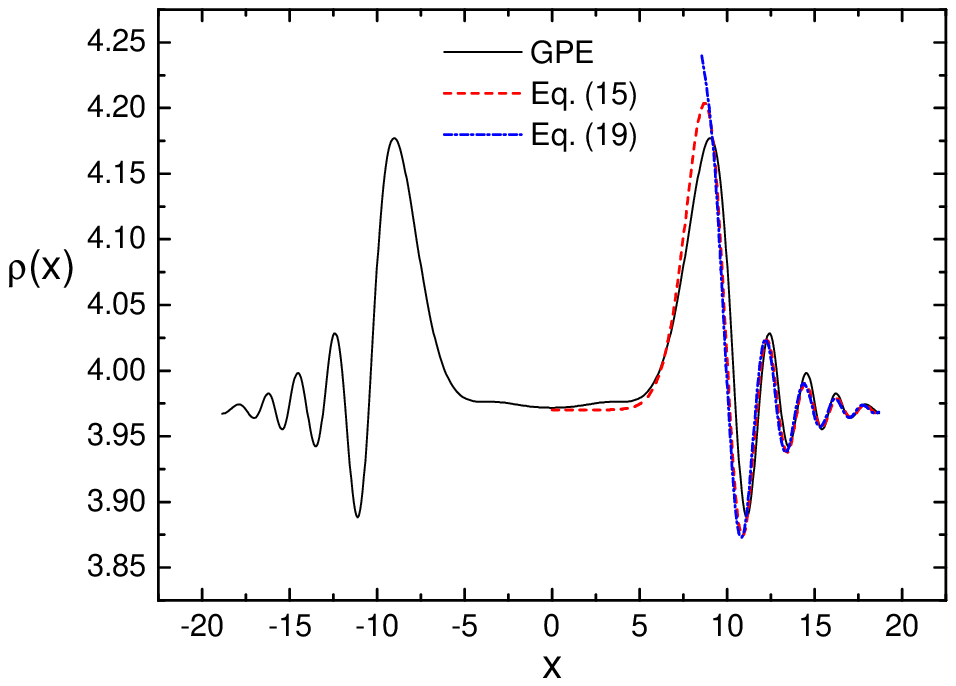}}
\caption{Left panel: Time evolution of the density disturbance in
the background BEC ($\rho(x) = |\psi(x)|^2$), when the
inter-species interaction constant is set to zero ($g_{12} = 0$)
at $t=0$, according to GP equation with $g_{22} = g = 1$. Right panel:
Comparison of the exact analytic solution (\ref{2-15}), asymptotic
approximation (\ref{2-19}), and numerical solution of the GP equation for
the density disturbance, corresponding to $t=4$ and $c=2$. The
initial disturbance has a gaussian form (\ref{2-21}) with
$a=0.95$. Space dependence for analytical solutions is shown for
$x>0$ at the moment $t=4$. Similar calculation gives a symmetric
distribution for $x<0$.}\label{fig2}
\end{figure}

Let us consider, for the sake of definiteness, the case of an
initial disturbance of density only, $\delta\phi_0=0$, so that
$W_1(k)=W_2(k)=\delta\rho_0(k)/(4\sqrt{\rho_0})$ and then
\begin{equation}\label{2-15}
\begin{array}{l}
    \delta\rho(x,t)=2\sqrt{\rho_0}A(x,t)=\frac1{4\pi}\int_{-\infty}^{\infty}
    \delta\rho_0(k)\left[e^{i(kx-\om(k)t)}+e^{i(kx+\om(k)t)}\right]dk\\
    =\frac1{\pi}\int_0^{\infty}\delta\rho_0(k)\cos(kx)\cos\left(tk\sqrt{c^2+\frac{k^2}4}
    \right)dk.
\end{array}
\end{equation}
The integrals here can be estimated for large $t$, as usual, by
the method of stationary phase. For $x>0$, only the first exponent
gives contribution to the integral,
\begin{equation}\label{2-16}
    \delta\rho(x,t)=\frac1{4\pi}\int_{-\infty}^{\infty}\delta\rho_0(k)
    e^{itf(k)}dk,\quad x>0,
\end{equation}
where
\begin{equation}\label{2-17}
    f(k)=k\left(\frac{x}t-\sqrt{c^2+\frac{k^2}4}\right).
\end{equation}
The values of $k$ at the points of stationary phase determined by the condition
$df/dk=0$ are equal to ($x>ct$)
\begin{equation}\label{2-18}
    k=\pm k_0(x,t),\quad k_0(x,t)=\frac{c}{\sqrt{2}}\left[\left(\frac{x}{ct}\right)^2
    -4+\frac{x}{ct}\sqrt{\left(\frac{x}{ct}\right)^2+8}\right]^{1/2},
\end{equation}
and integration yields
\begin{equation}\label{2-19}
    \delta\rho(x,t)=\frac{\delta\rho_0(k)}{\sqrt{2\pi t\left|\frac{d^2f}{dk^2}\right|_0}}
    \cos\left[tf(k_0)-\frac{\pi}4\right]
\end{equation}
where
\begin{equation}\label{2-20}
    \left|\frac{d^2f}{dk^2}\right|_0=\frac{k_0(6c^2+k_0^2)}{(4c^2+k_0^2)^{3/2}}.
\end{equation}
Here $k_0$ is a function of $x$ and $t$ determined by
Eq.~(\ref{2-18}). In Fig.~\ref{fig2} we compare the exact solution
(\ref{2-15}) corresponding to a gaussian initial disturbance
\begin{equation}\label{2-21}
    \delta\rho_0(x)=\frac1{\sqrt{\pi}a}\exp\left(-\frac{x^2}{a^2}\right),\quad
    \delta\rho_0(k)=\exp\left(-\frac{k^2a^2}4\right),
\end{equation}
with its approximation (\ref{2-19}), and corroborate these two
curves with numerical solution of the GP equation where we introduce as
initial condition the wave profiles shown in the left panel of
Fig.~\ref{fig1}.
As we see, the asymptotic formula (\ref{2-19}) is quite accurate
even for not very large values of $t$ along almost all wave packet
where $|x|>ct$.

\section{Two-dimensional condensates}

\subsection{Decay of small disturbances to wave packets}

This problem can be solved by the same method which was used in
1D case with replacement $\prt^2/\prt x^2 \to \Delta_\bot\equiv
\prt^2/\prt x^2 +\prt^2/\prt y^2$. As a result we obtain the general
solution in the form
\begin{equation}\label{3-1}
    A({\mathbf r},t)=\int W_1(\bk)e^{i(\bk\bfr-\om(k)t)}\frac{d\bk}{(2\pi)^2}+
    \int W_2(\bk)e^{i(\bk\bfr+\om(k)t)}\frac{d\bk}{(2\pi)^2},
\end{equation}
where $W_{1,2}(\bk)$ are determined again from the initial
conditions, which gives
\begin{equation}\label{3-2}
\begin{array}{l}
    W_1(\bk)=\frac1{4\sqrt{\rho_0}}\left(\delta\rho_0(\bk)+i\frac{k^2\rho_0}{\om(k)}
    \delta\phi_0(\bk)\right)\\
    W_2(\bk)=\frac1{4\sqrt{\rho_0}}\left(\delta\rho_0(\bk)-i\frac{k^2\rho_0}{\om(k)}
    \delta\phi_0(\bk)\right).
    \end{array}
\end{equation}
For cylindrically symmetric initial disturbance of density only ($\delta\phi_0(\bk)=0$)
we get for the wave of density $\delta\rho(r,t)=2\sqrt{\rho_0}\,A(r,t)$ the expressions
similar to Eqs.~(\ref{2-15}),
\begin{equation}\label{3-3}
\begin{array}{l}
    \delta\rho(r,t)=\frac1{8\pi^2}\int_0^{2\pi}\int_0^{\infty}
    \delta\rho_0(k)\left[e^{i(kr\cos\vfi-\om(k)t)}+e^{i(kr\cos\vfi+\om(k)t)}\right]kdkd\vfi,\\
    =\frac1{2\pi}\int_0^{\infty}\delta\rho_0(k)\cos\left(tk\sqrt{c^2+\frac{k^2}4}\right)
    J_0(kr)kdk,
\end{array}
\end{equation}
where
\begin{equation}\label{3-4}
    J_0(z)=\frac1{2\pi}\int_0^{\infty}e^{iz\cos\vfi}d\vfi
\end{equation}
is the Bessel function of the first kind.

For $kr\gg1$ the integral in (\ref{3-3}) can be estimated by the method
of stationary phase with the use of asymptotic expression for the Bessel function,
\begin{equation}\label{3-5}
    J_0(kr)\approx\sqrt{\frac{2}{\pi kr}}\cos(kr-\pi/4).
\end{equation}
As a result we get
\begin{equation}\label{3-6}
    \delta\rho(r,t)=\frac{\delta\rho_0(k)\sqrt{k_0}}{2\pi\sqrt{r t\left|\frac{d^2f}{dk^2}\right|_0}}
    \sin\left[tf(k_0)\right],
\end{equation}
where
\begin{equation}\label{3-8}
    k_0(r,t)=\frac{c}{\sqrt{2}}\left[\left(\frac{r}{ct}\right)^2
    -4+\frac{r}{ct}\sqrt{\left(\frac{r}{ct}\right)^2+8}\right]^{1/2},\quad r>ct,
\end{equation}
\begin{equation}\label{3-9}
    f(k)=k\left(\frac{r}t-\sqrt{c^2+\frac{k^2}4}\right),
\end{equation}
and $|d^2f/dk^2|_0$ is given by (\ref{2-20}).

\begin{figure}[htb]
\centerline{
\includegraphics[width=6cm, height=6cm,clip]{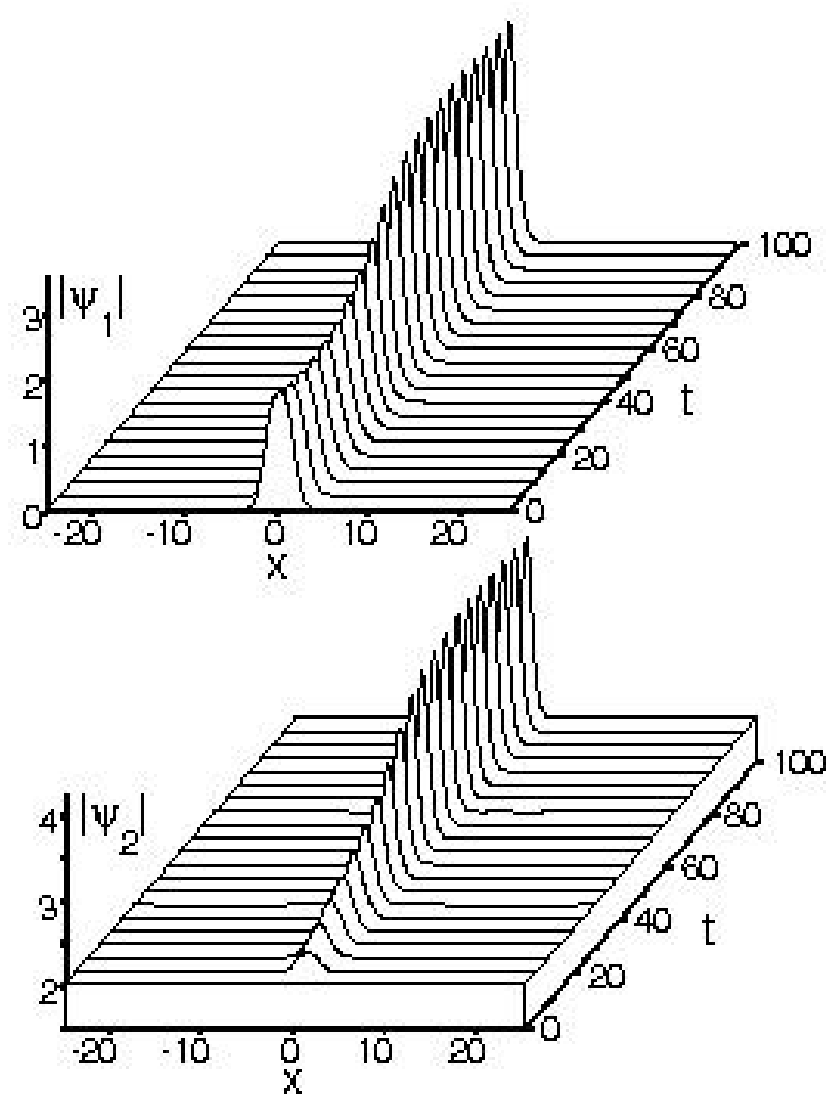} \qquad
\includegraphics[width=6cm, height=6cm,clip]{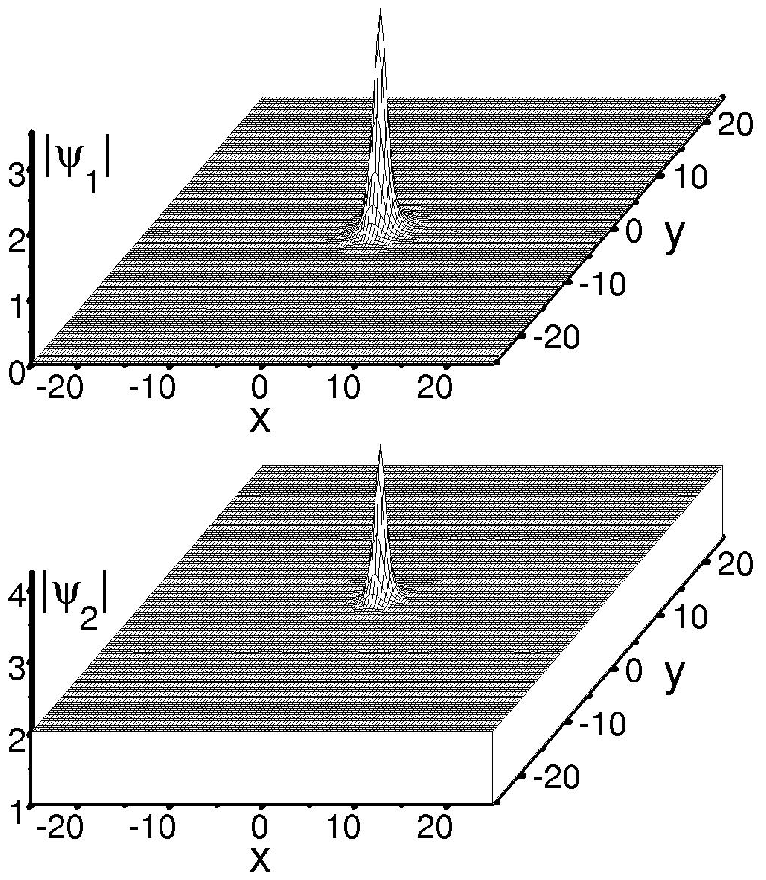}}
\caption{Left panel: Formation of a density hump on the background
condensate $\psi_2(x,y)$, shown at $y=0$ cross section, when the
inter-species interaction constant is slowly raised from zero
according to $g_{12} = -0.5 \mathrm{tanh}(5t/t_{end})$, with
$t_{end}=100$. Due to inter-species attraction, the drop-like
repulsive condensate $\psi_1(x,y)$ with a norm $N=6\pi$ becomes
more narrow. Right panel: The final stationary coupled state of
the two repulsive BEC's with $g_{11} = 1$, $g_{22} = 1$, with
attractive inter-species interaction constant $g_{12} = -0.5$.}
\label{fig3}
\end{figure}

\begin{figure}[htb]
\centerline{
\includegraphics[width=6cm, height=6cm,clip]{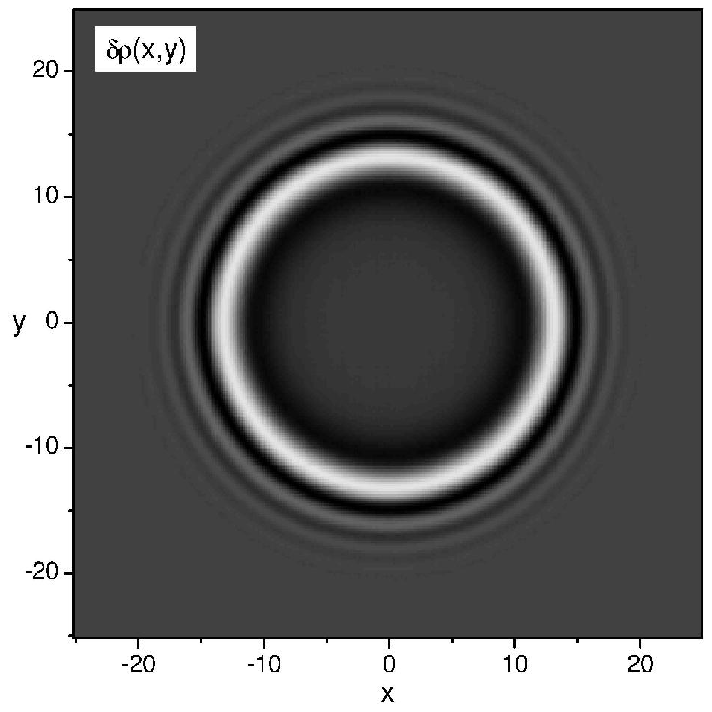} \qquad
\includegraphics[width=6cm, height=6cm,clip]{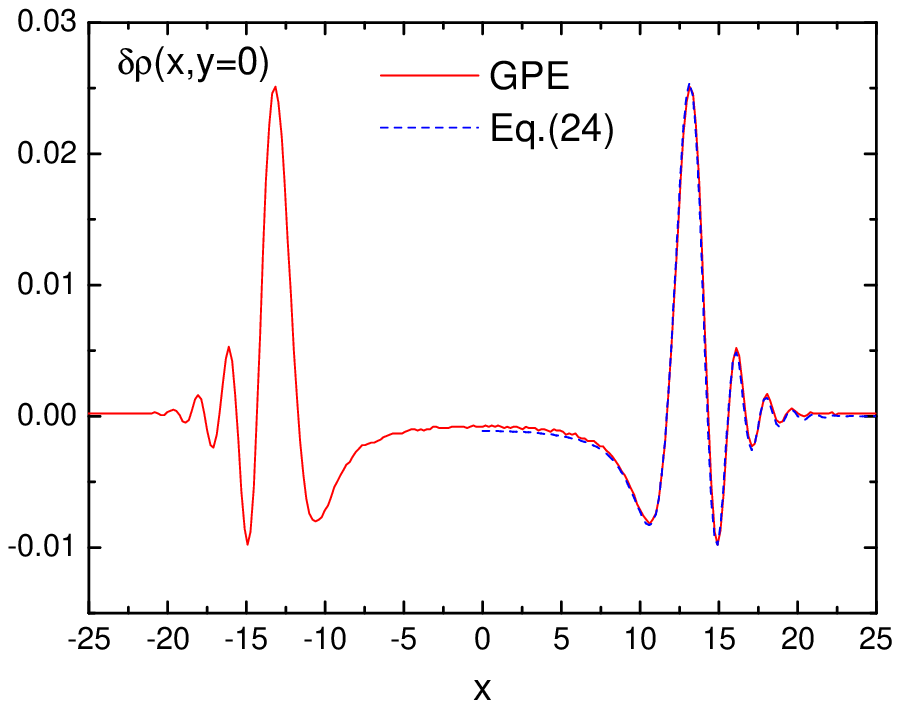}}
\caption{Left panel: Density plot $\rho(r)=|\psi(x,y)|^2$ of the
repulsive background condensate at $t=4$ after turning off the
inter-species interaction constant $g_{12} = 0$, according to
numerical solution of the GP equation. White stripes correspond to
higher density. Right panel: Comparison between analytical
expression (24), for the density disturbance in the background
condensate at $t=4$ with corresponding numerical solution of the
GP equation ($|\psi(x,y=0)|^2-\rho_0$), shown for $y=0$ cross
section. Parameters: $\rho_0=9$, $c=3$, $a=1.0$. } \label{fig4}
\end{figure}

In order to compare the analytical predictions with numerical
solution of the GP equation we need to perform a sequence of
simulations as in 1D case. However, there is an essential
difference in the 2D case due to the fact that no stable localized
solution for the drop-like condensate exists for the 2D GP
equation without an external potential. To this regard we recall
that the attractive 2D NLS equation admits a localized solution,
the so called Townes soliton, which exist only for  a critical
value of the norm (number of atoms) and is unstable against small
fluctuations of the critical norm (overcritical norms lead to
collapse while undercritical ones lead to background decays). To
avoid this problem, we prepare the drop-like condensate in the
ground state of the parabolic trap $V(x,y) = v_0 (x^2 + y^2)$ with
strength $v_0=1$, using the imaginary-time relaxation method
\cite{chiofalo}. Then we introduce the obtained ground state
solution for the drop-like condensate $\psi_1(x,y)$ and constant
amplitude solution for the background condensate
($\psi_2(x,y)=2.0$) as initial condition in the coupled GP
equation with $g_{12}=0$, and slowly increase the inter-species
interaction constant to a finite value $g_{12}=-0.5$ (see
Fig.~\ref{fig3} left panel). It is remarkable that after the
localized coupled state has been created by this method, it
remains stable even after the parabolic potential is removed, due
to the stabilizing effect induced by the interspecies coupling
(this state exists only due to coupling). As a result the initial
state shown in the right panel of Fig.~\ref{fig3} is obtained. The
density waves in the background condensate can be generated by
fast turning off the interaction constant $g_{12} \rightarrow 0$.
Decaying density hump induces circular wave in the background BEC
traveling outward from the center. We have shown in
Fig.~\ref{fig4} the dependence of $\delta\rho$ on $r$ at fixed
value of time ($t=4$) for the gaussian initial distribution
\begin{equation}\label{3-10}
    \delta\rho_0(r)=\frac1{{\pi}a^2}\exp\left(-\frac{r^2}{a^2}\right),\quad
    \delta\rho_0(k)=\exp\left(-\frac{k^2a^2}4\right),
\end{equation}
with $a=1$ calculated according to the exact expression
(\ref{3-3}). The asymptotic approximation (\ref{3-6}) agrees very
well with the exact expression (\ref{3-3}) at $x>ct$ (not shown in
figure because the corresponding lines overlap).

\section{Two-component condensates}
Let us now consider the relevant case of a homogeneous binary BEC
mixture, when both components have a finite background.
Two-component condensates are described in our non-dimensional
variables by the equations
\begin{equation}\label{4-1}
    \begin{array}{l}
    i \frac{\partial \psi_1}{\partial t} +\frac{1}{2}\Delta \psi_1 +
[\mu_1-(g_{11}\left| \psi_1 \right|^2 + g_{12}\left| \psi_2 \right|^2)]\psi_1 =0 ,\\
i \frac{\partial \psi_2}{\partial t}+\frac{1}{2}\Delta\psi_2+
[\mu_2-(g_{12}\left| \psi_1 \right|^2 + g_{22}\left| \psi_2 \right|^2)]\psi_2 =0 ,
    \end{array}
\end{equation}
where
\begin{equation}\label{4-2}
    \mu_1=g_{11}\rho_{10}+g_{12}\rho_{20},\quad \mu_2=g_{12}\rho_{10}+g_{22}\rho_{20}
\end{equation}
are chemical potentials of two species with equilibrium uniform densities
$\rho_{10}$ and $\rho_{20}$; we keep here the general notation for the
Laplacian $\Delta$ which will be specified later depending on the number
of spatial dimensions in the problem under consideration.

In disturbed condensates we have $\psi_1=\sqrt{\rho_{10}}+\delta\psi_1,$
$\psi_2=\sqrt{\rho_{20}}+\delta\psi_2$, where evolution of disturbances
is governed by the linearized equations
\begin{equation}\label{4-3}
    \begin{array}{l}
    i\delta\psi_1+\frac12\Delta\delta\psi_1-g_{11}\rho_{10}(\delta\psi_1+
    \delta\psi_1^*)-g_{12}\sqrt{\rho_{10}\rho_{20}}(\delta\psi_2+\delta\psi_2^*)=0,\\
    i\delta\psi_2+\frac12\Delta\delta\psi_2-g_{12}\sqrt{\rho_{10}\rho_{20}}
    (\delta\psi_1+\delta\psi_1^*)-g_{22}\rho_{20}(\delta\psi_2+\delta\psi_2^*)=0.
    \end{array}
\end{equation}
As before, we introduce
\begin{equation}\label{4-4}
    \delta\psi_1=A_1+iB_1,\quad \delta\psi_2=A_2+iB_2,
\end{equation}
so that after substitution into (\ref{4-3}), separation of real and imaginary
parts followed by exclusion of $B_1$ and $B_2$ we arrive at the system
\begin{equation}\label{4-5}
    \begin{array}{l}
    A_{1,tt}-g_{11}\rho_{10}\Delta A_1+\frac14\Delta^2A_1-g_{12}\sqrt{\rho_{10}\rho_{20}}
    \Delta A_2=0,\\
    A_{2,tt}-g_{22}\rho_{20}\Delta A_2+\frac14\Delta^2A_2
    -g_{12}\sqrt{\rho_{10}\rho_{20}}\Delta A_1=0.
    \end{array}
\end{equation}
For plane waves $A_1,\,A_2\propto\exp[i(\bk\bfr-\om t)]$ we
reproduce the known dispersion laws
\begin{equation}\label{4-6}
    \om_{\pm}=k\sqrt{c_{\pm}^2+\frac{k^2}4},
\end{equation}
where the sound velocities corresponding to infinite wavelengths are equal to
\begin{equation}\label{4-7}
    c_{\pm}=\frac1{\sqrt{2}}\left[g_{11}\rho_{10}+g_{22}\rho_{20}
    \pm\sqrt{(g_{11}\rho_{10}-g_{22}\rho_{20})^2+4g_{12}^2\rho_{10}\rho_{20}}\right]^{1/2}.
\end{equation}
The ratios $A_1/A_2$ in these two modes of linear waves are given by
\begin{equation}\label{4-8}
    \frac{A_1}{A_2}=\frac{2g_{12}\sqrt{\rho_{10}\rho_{20}}}
    {g_{22}\rho_{20}-g_{11}\rho_{10}\pm(c_+^2-c_-^2)}.
\end{equation}
Thus, the general solution of the system (\ref{4-5}) is given by
\begin{equation}\label{4-9}
\begin{array}{l}
    A_1(\bfr,t)=\int W_{11}^+(\bk)e^{i(\bk\bfr-\om_+t)}\frac{d^Dk}{(2\pi)^D}
    +\int W_{12}^+(\bk)e^{i(\bk\bfr+\om_+t)}\frac{d^Dk}{(2\pi)^D}\\
    +\int W_{11}^-(\bk)e^{i(\bk\bfr-\om_-t)}\frac{d^Dk}{(2\pi)^D}
    +\int W_{12}^-(\bk)e^{i(\bk\bfr+\om_-t)}\frac{d^Dk}{(2\pi)^D},
\end{array}
\end{equation}
\begin{equation}\label{4-10}
\begin{array}{l}
    A_2(\bfr,t)=\int W_{21}^+(\bk)e^{i(\bk\bfr-\om_+t)}\frac{d^Dk}{(2\pi)^D}
    +\int W_{22}^+(\bk)e^{i(\bk\bfr+\om_+t)}\frac{d^Dk}{(2\pi)^D}\\
    +\int W_{21}^-(\bk)e^{i(\bk\bfr-\om_-t)}\frac{d^Dk}{(2\pi)^D}
    +\int W_{22}^-(\bk)e^{i(\bk\bfr+\om_-t)}\frac{d^Dk}{(2\pi)^D},
\end{array}
\end{equation}
where $D$ denotes dimension of the space. According to (\ref{4-8}) the
Fourier components are related with each other by the equations
\begin{equation}\label{4-11}
    \frac{W_{11}^{\pm}}{W_{21}^{\pm}}=\frac{W_{12}^{\pm}}{W_{22}^{\pm}}=
    \frac{2g_{12}\sqrt{\rho_{10}\rho_{20}}}
    {g_{22}\rho_{20}-g_{11}\rho_{10}\pm(c_+^2-c_-^2)},
\end{equation}
that is only four of them are independent of each other and should
be determined from the initial conditions.

To simplify the notation, in what follows we shall consider the
case when there exist only initial disturbances
$\delta\rho_{10}(\bfr),\delta\rho_{20}(\bfr)$ of the densities.
Then
\begin{equation}\label{4-12}
    W_{11}^{\pm}=W_{12}^{\pm}\equiv W_{1}^{\pm},\quad
    W_{21}^{\pm}=W_{22}^{\pm}\equiv W_{2}^{\pm},
\end{equation}
and these four functions must satisfy the linear system
\begin{equation}\label{4-13}
    \begin{array}{l}
    A_{10}(\bk)=2(W_1^++W_1^-),\quad A_{20}(\bk)=2(W_2^++W_2^-),\\
    \frac{W_{1}^{+}}{W_{2}^{+}}=
    \frac{2g_{12}\sqrt{\rho_{10}\rho_{20}}}
    {g_{22}\rho_{20}-g_{11}\rho_{10}+(c_+^2-c_-^2)},\quad
    \frac{W_{1}^{-}}{W_{2}^{-}}=
    \frac{2g_{12}\sqrt{\rho_{10}\rho_{20}}}
    {g_{22}\rho_{20}-g_{11}\rho_{10}-(c_+^2-c_-^2)},
    \end{array}
\end{equation}
which can be readily solved. Taking into account that
\begin{equation}\label{4-14}
    A_{10}(\bk)=\frac1{2\sqrt{\rho_{10}}}\delta\rho_{10}(\bk),\quad
    A_{20}(\bk)=\frac1{2\sqrt{\rho_{20}}}\delta\rho_{20}(\bk),
\end{equation}
we can find the waves of the densities of the two species.

In 1D case the final formulae read
\begin{equation}\label{4-15}
    \begin{array}{l}
    \delta\rho_1(x,t)=\frac1{2\pi}\int_0^{\infty}\delta\rho_1(k)\cos kx
    \Bigg[\left(1+\frac{g_{11}\rho_{10}-g_{22}\rho_{20}}{c_+^2-c_-^2}\right)
    \cos\left(tk\sqrt{c_+^2+\frac{k^2}4}\right)\\
    +\left(1-\frac{g_{11}\rho_{10}-g_{22}\rho_{20}}{c_+^2-c_-^2}\right)
    \cos\left(tk\sqrt{c_-^2+\frac{k^2}4}\right)\Bigg]dk\\
    +\frac{g_{12}\rho_{10}}{\pi(c_+^2-c_-^2)}\int_0^{\infty}\delta\rho_2(k)\cos kx
    \left[\cos\left(tk\sqrt{c_+^2+\frac{k^2}4}\right)-
    \cos\left(tk\sqrt{c_-^2+\frac{k^2}4}\right)\right]dk,
    \end{array}
\end{equation}
\begin{equation}\label{4-16}
    \begin{array}{l}
    \delta\rho_2(x,t)=\frac1{2\pi}\int_0^{\infty}\delta\rho_2(k)\cos kx
    \Bigg[\left(1-\frac{g_{11}\rho_{10}-g_{22}\rho_{20}}{c_+^2-c_-^2}\right)
    \cos\left(tk\sqrt{c_+^2+\frac{k^2}4}\right)\\
    +\left(1+\frac{g_{11}\rho_{10}-g_{22}\rho_{20}}{c_+^2-c_-^2}\right)
    \cos\left(tk\sqrt{c_-^2+\frac{k^2}4}\right)\Bigg]dk\\
    +\frac{g_{12}\rho_{20}}{\pi(c_+^2-c_-^2)}\int_0^{\infty}\delta\rho_1(k)\cos kx
    \left[\cos\left(tk\sqrt{c_+^2+\frac{k^2}4}\right)-
    \cos\left(tk\sqrt{c_-^2+\frac{k^2}4}\right)\right]dk.
    \end{array}
\end{equation}
Similar formulae are obtained for two-dimensional cylindrically
symmetric waves as:
\begin{equation}\label{4-17}
    \begin{array}{l}
    \delta\rho_1(r,t)=\frac1{4\pi}\int_0^{\infty}\delta\rho_1(k)J_0(kr)
    \Bigg[\left(1+\frac{g_{11}\rho_{10}-g_{22}\rho_{20}}{c_+^2-c_-^2}\right)
    \cos\left(tk\sqrt{c_+^2+\frac{k^2}4}\right)\\
    +\left(1-\frac{g_{11}\rho_{10}-g_{22}\rho_{20}}{c_+^2-c_-^2}\right)
    \cos\left(tk\sqrt{c_-^2+\frac{k^2}4}\right)\Bigg]kdk\\
    +\frac{g_{12}\rho_{10}}{2\pi(c_+^2-c_-^2)}\int_0^{\infty}\delta\rho_2(k)J_0(kr)
    \left[\cos\left(tk\sqrt{c_+^2+\frac{k^2}4}\right)-
    \cos\left(tk\sqrt{c_-^2+\frac{k^2}4}\right)\right]kdk,
    \end{array}
\end{equation}
\begin{equation}\label{4-18}
    \begin{array}{l}
    \delta\rho_2(r,t)=\frac1{4\pi}\int_0^{\infty}\delta\rho_2(k)J_0(kr)
    \Bigg[\left(1-\frac{g_{11}\rho_{10}-g_{22}\rho_{20}}{c_+^2-c_-^2}\right)
    \cos\left(tk\sqrt{c_+^2+\frac{k^2}4}\right)\\
    +\left(1+\frac{g_{11}\rho_{10}-g_{22}\rho_{20}}{c_+^2-c_-^2}\right)
    \cos\left(tk\sqrt{c_-^2+\frac{k^2}4}\right)\Bigg]kdk\\
    +\frac{g_{12}\rho_{10}}{2\pi(c_+^2-c_-^2)}\int_0^{\infty}\delta\rho_1(k)J_0(kr)
    \left[\cos\left(tk\sqrt{c_+^2+\frac{k^2}4}\right)-
    \cos\left(tk\sqrt{c_-^2+\frac{k^2}4}\right)\right]kdk.
    \end{array}
\end{equation}
It is easy to see that if there is no interaction between two
components of BECs, then these formulae reduce to the already
known results for one-component condensates.

Integrals in (\ref{4-15})-(\ref{4-18}) can be estimated by the
method of stationary phase which gives the approximate expressions
for disturbances of the densities in the regions $x>c_+t$ or
$r>c_+t$ where we can observe the interference pattern of two
modes.

\begin{figure}[htb]
\centerline{\includegraphics[width=8cm,height=8cm,clip]{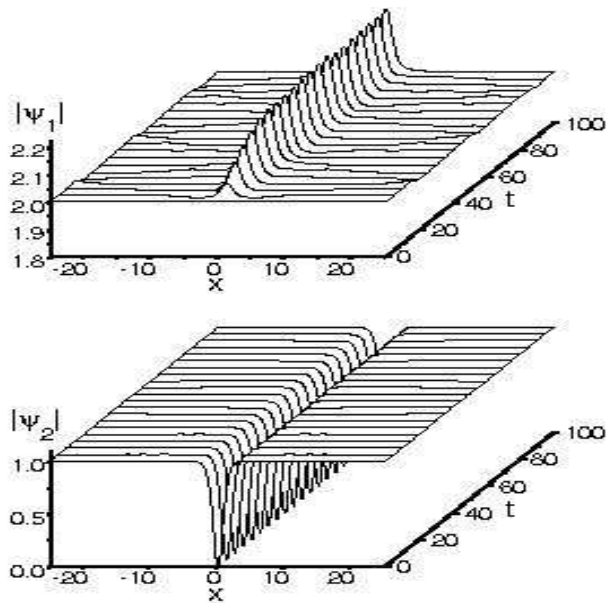}}
\caption{Formation of a density defect in two coupled 1D repulsive
BEC's with initial wave amplitudes $\psi_1(x)=2,
\psi_2(x)=\mathrm{tanh}(x)$, and $g_{11} = 1, g_{22} = 1$, when
the repulsive inter-species interaction is slowly turned on
according to law $g_{12} = 0.5 \, \mathrm{tanh}(5 t/t_{end})$,
with $t_{end}=100$. Dark soliton in one component gives rise to a
density hump in the other component.} \label{fig5}
\end{figure}

The present setting with homogeneously mixed two repulsive BECs
corresponds to the most relevant experimental situation. The case
of two repulsive 1D BECs is akin to previously discussed one (see
Fig.~\ref{fig1}), if there is a dark soliton in one component.
When the repulsive inter-species interaction is slowly turned on a
density hump emerges in the other component, as illustrated in
Fig.~\ref{fig5}. Linear waves then can be generated by a fast
turning off the inter-species interaction.

Another approach for generation of density disturbances in coupled
repulsive BEC's involves local change of the intra- or
inter-species interaction constants by a tightly focused laser
beam via optically induced Feshbach resonance \cite{fatemi}.
Stationary bright-bright, dark-dark and dark-bright coupled
localized states on finite backgrounds can emerge in mixture BEC's
when the interaction constants are locally changed. In particular,
a stationary bright-bright (dark-dark) localized state in all
repulsive case occurs if the inter-species repulsion is locally
reduced (increased). A bright-dark localized state emerges when
the intra-species repulsion is locally reduced in one of the
components.

\begin{figure}[htb]
\centerline{
\includegraphics[width=6cm,height=4cm,clip]{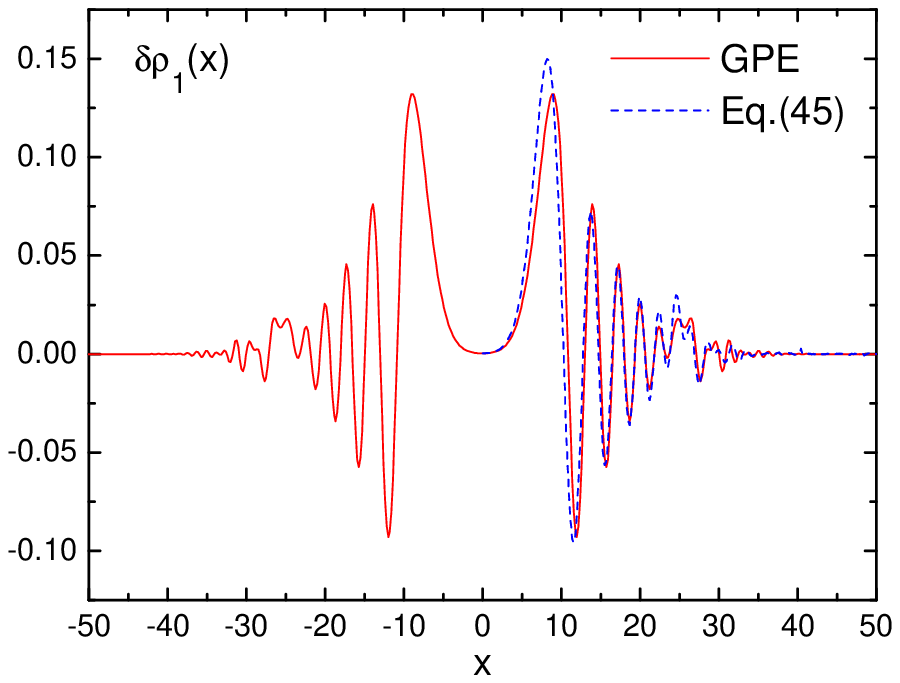} \quad
\includegraphics[width=6cm,height=4cm,clip]{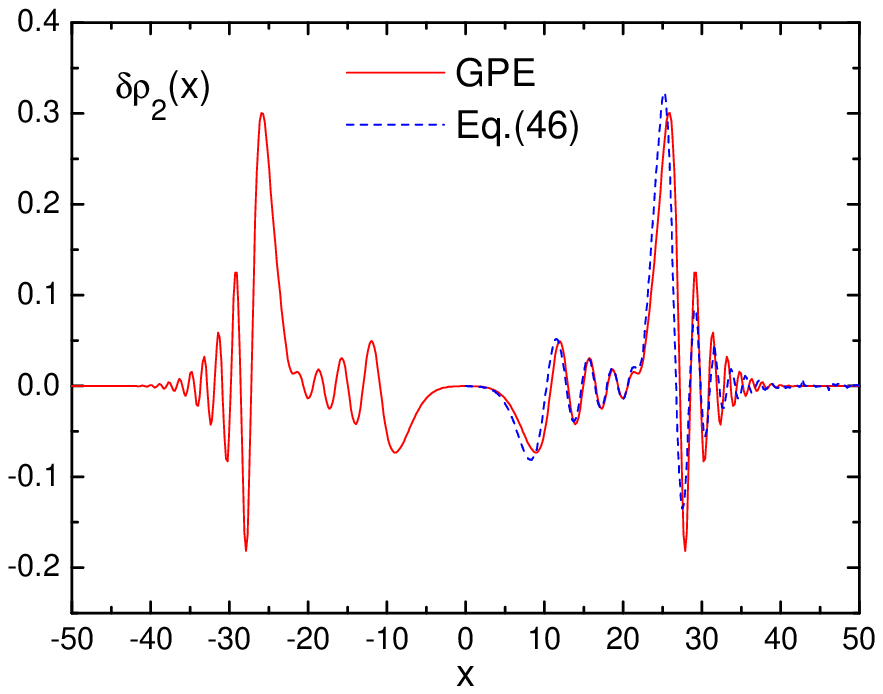}}
\caption{Density modulations at $t=8$ according to Eqs. (45)-(46)
and numerical solution of the coupled GPE. Parameters: $\rho_{10}
=1$, $\rho_{20} = 9$, $a=1$, $g_{11}=1$, $g_{22}=1$,
$g_{12}=0.5$.} \label{fig6}
\end{figure}

After the density defect (hump or hole) has been created, a fast
change of the inter-species interaction coefficient leads to
generation of waves in the condensate caused by decaying density
disturbance. These results are illustrated in Fig.~\ref{fig6},
where comparison between the analytical prediction and numerical
simulation of the GPE is also presented.

Since there is no stable localized solution (like dark soliton in
1D repulsive case) of the GP equation in higher dimensional
settings, creation of density disturbance in this case also can be
done by illuminating the homogeneous mixture BEC by tightly
focused laser beam designed to change the intra- or inter-species
interaction constant via optically induced Feshbach resonance.

\begin{figure}[htb]
\centerline{
\includegraphics[width=6cm,height=4cm,clip]{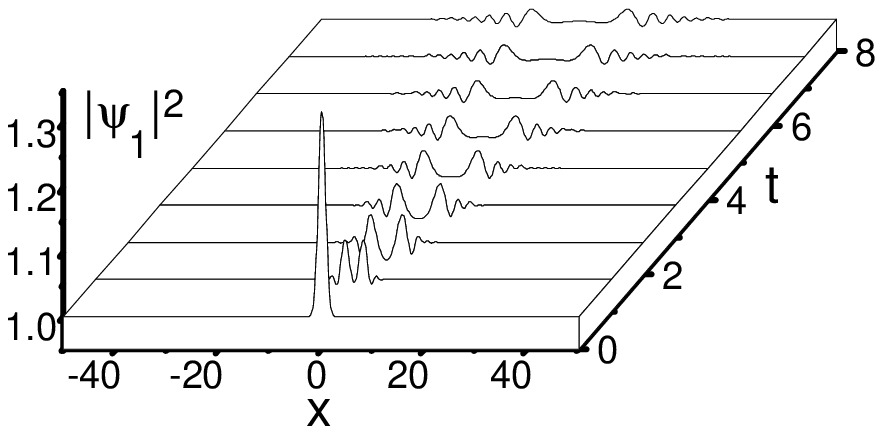} \quad
\includegraphics[width=6cm,height=4cm,clip]{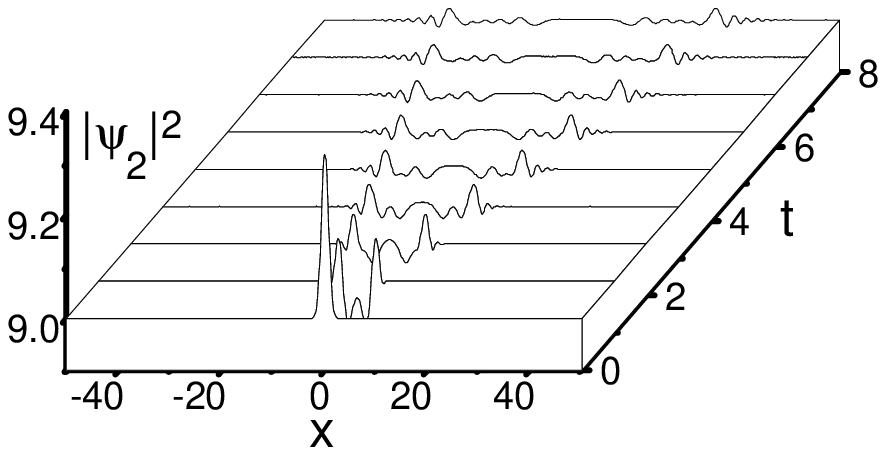}}
\caption{Evolution of waves in a 2D mixture BEC emerged from the
decay of a Gaussian density disturbance with $a=1$ in both
components shown at $y=0$ cross section according to numerical
solution of the coupled GP equations. Background wave amplitudes
of components and nonlinear coefficients are equal to
$\psi_1(x,y)=1$, $\psi_2(x,y)=3$, $g_{11}=1$, $g_{22}=1$,
$g_{12}=0.5$. Due to different sound velocities in the components
radially outward travelling waves quickly separate.} \label{fig7}
\end{figure}

\begin{figure}[htb]
\centerline{
\includegraphics[width=6cm,height=6cm,clip]{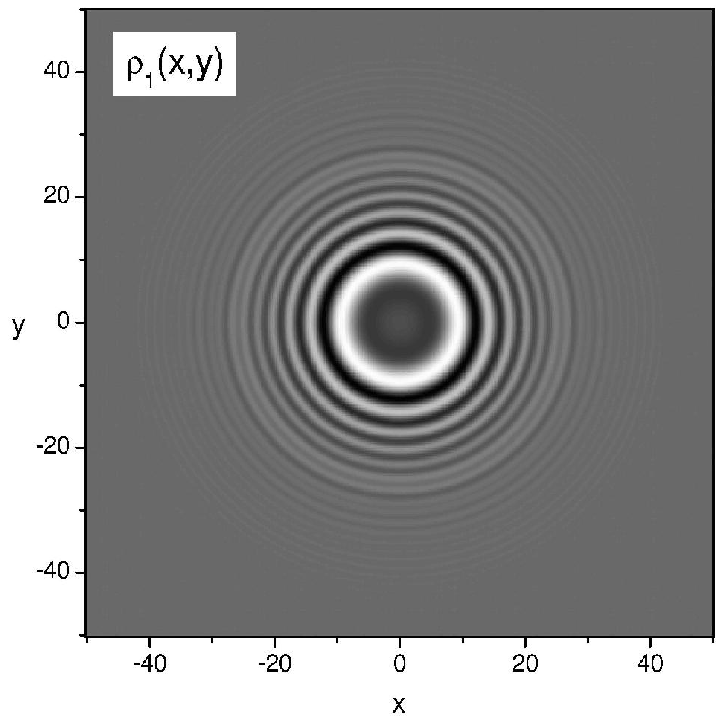} \quad
\includegraphics[width=6cm,height=6cm,clip]{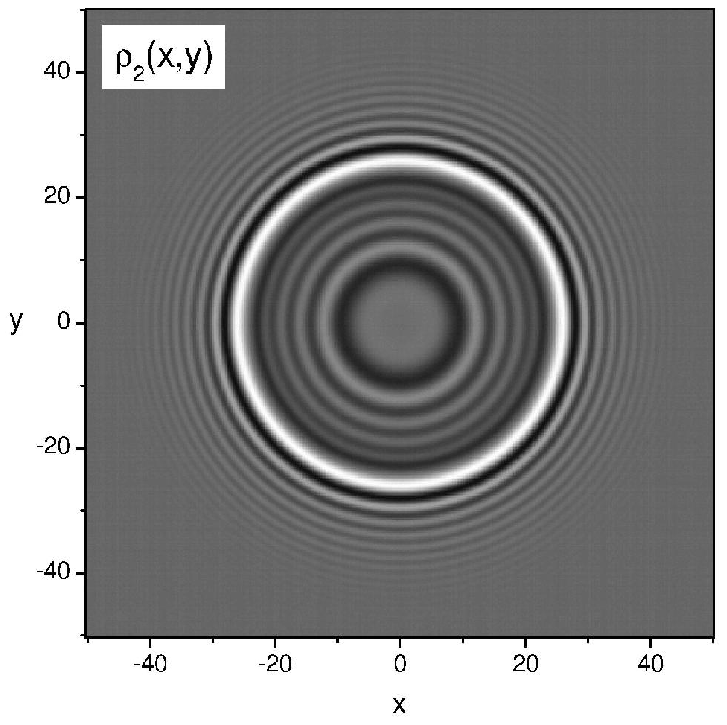}}
\caption{Density plots of two components corresponding to the last
time section at $t=8$ of the previous figure \ref{fig7}. In white
areas the density is higher. Separation of waves in the two
components due to different sound velocities is clearly seen.}
\label{fig8}
\end{figure}

\begin{figure}[htb]
\centerline{
\includegraphics[width=6cm,height=4cm,clip]{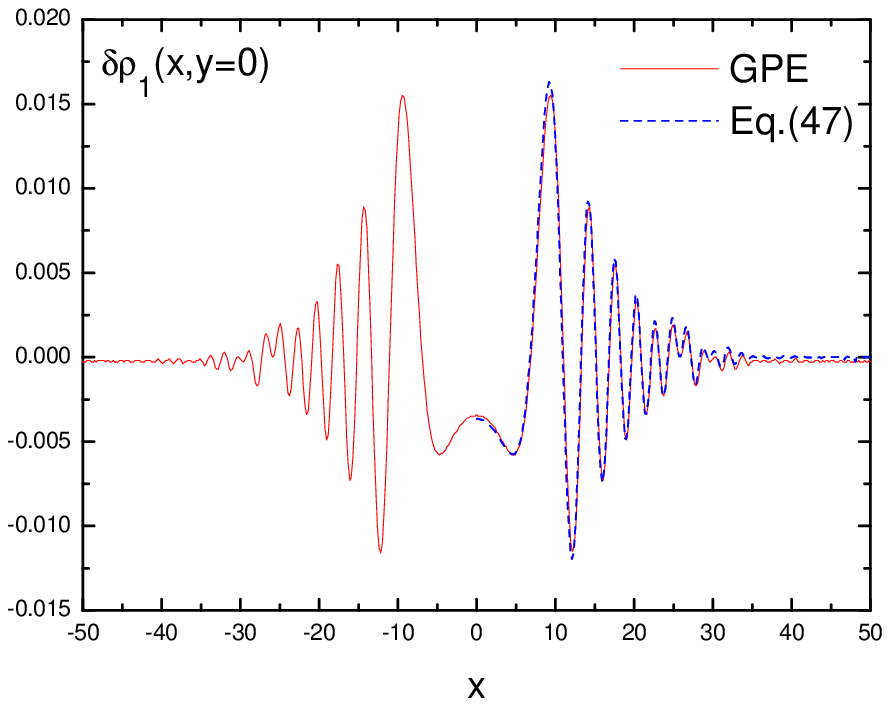} \quad
\includegraphics[width=6cm,height=4cm,clip]{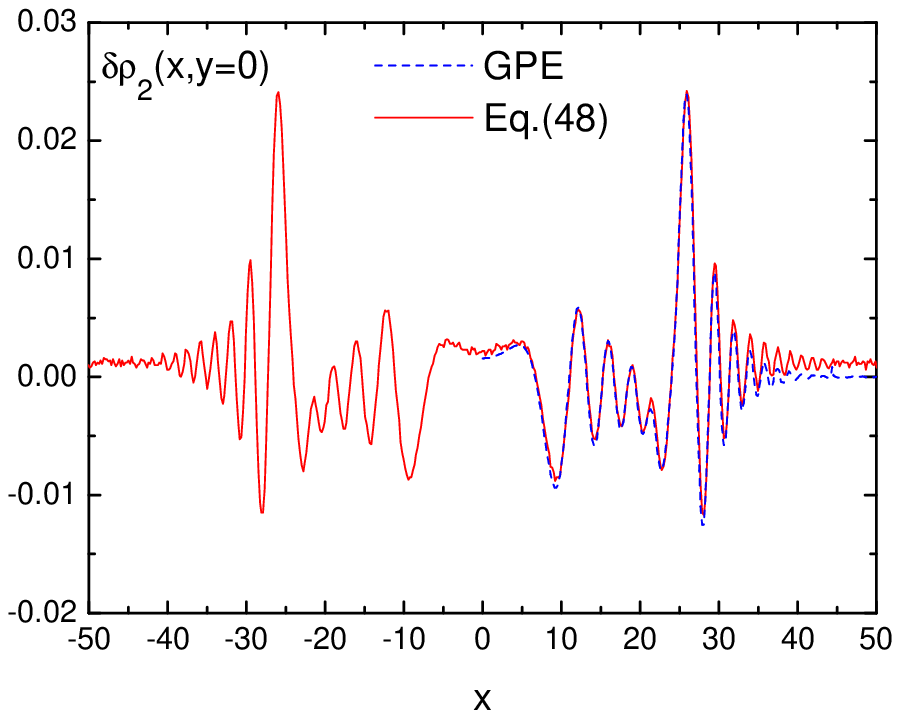}}
\caption{Comparison of density distributions provided by Eqs. (47)
and (48) with numerical solution of the coupled GP equation for
the same parameters as in Fig. \ref{fig7} for $t=8$. Relative
positions of the wave maximums in the second and first modes
$26/9 \simeq 3$ well match the ratio of corresponding sound velocities
$c_{+}/c_{-}$ which have the same order of magnitude as
group velocities of our wave packets built of harmonics with
wavelengths about healing length.}\label{fig9}
\end{figure}

In Figs.~\ref{fig7}--\ref{fig9} the generation, further evolution
and pattern of linear waves at particular time are depicted for
the 2D coupled repulsive BEC's. As can be seen from Figs.~\ref{fig4},
\ref{fig6} and \ref{fig9}, the agreement between
analytical predictions and numerical solution of the GPE is quite
good.


\section{Conclusions}

We have investigated sound waves in two-component BECs and
proposed a new method of wave generation which is based on a fast
change of the inter-species interaction constant. To illustrate
the idea, we have performed numerical simulations of the creation
of a density defect (hump or hole) on the background condensate
and its evolution into matter wavepackets after the fast change of
the interspecies interaction constant. This method of wave
generation was used to investigate both numerically and
analytically sounds waves arising from the release of  a drop-like
condensate immersed into a second large repulsive condensate and
from homogeneous mixture of two repulsive BEC's both with finite
backgrounds. We demonstrated that propagation of linear waves
created by this approach is well described by the Gross-Pitaevskii
equation. Explicit analytical formulae were obtained for the
space-time evolution of density waves in coupled BECs, starting
from the linearized GP equations. The comparison of the analytical
results with direct numerical simulations of the coupled GPE
equations showed an excellent agreement.

Although in this paper we have restricted our considerations  to
weak density disturbances (e.g. amplitude of the disturbance is
small compared to the amplitude of the background condensate), the
proposed method can be used to investigate also ship waves and
dispersive shocks in binary BEC mixtures as it will be discussed
elsewhere.

\ack BBB acknowledges partial support from the Fund for
Fundamental Research of the Uzbek Academy of Sciences under Grant
No.~10-08. AMK wishes to thank the Department of Physics ``E.R.
Caianiello'' of the University of Salerno, where part of this work
was done, for the hospitality received and for financial support.

\end{document}